\documentclass[twocolumn,english,aps,prb,twocolum,superscriptaddress,nobibnotes,amsmath,amssymb,floatfix]{revtex4-1}

\usepackage[utf8]{inputenc}
\usepackage[english]{babel}
\usepackage{amsmath,amsfonts,amssymb}
\usepackage{upgreek}
\usepackage[T1]{fontenc}
\usepackage{url}
\usepackage{placeins}

\usepackage[version=3]{mhchem}

\usepackage{epstopdf}
\usepackage{graphicx}

\usepackage{titlesec}
\titleformat{\section}{\normalfont\large\bfseries\filcenter}{\thesection}{1em}{}
\titlespacing{\section}{0pt}{10pt}{5pt}

\begin{document}

\title{Narrow linewidth lasing and soliton Kerr-microcombs with ordinary laser diodes}

\author{N.~G.~Pavlov}
\affiliation{Russian Quantum Center, 143025, Skolkovo, Russia}
\affiliation{Moscow Institute of Physics and Technology, 141700, Dolgoprudny, Russia}

\author{S.~Koptyaev}
\affiliation{Samsung R\&D Institute Russia, SAIT-Russia Laboratory, 127018, Moscow, Russia}

\author{G.~Lihachev}
\affiliation{Russian Quantum Center, 143025, Skolkovo, Russia}
\affiliation{Faculty of Physics, M.V. Lomonosov Moscow State University, 119991
Moscow, Russia}

\author{A.~S.~Voloshin}
\affiliation{Russian Quantum Center, 143025, Skolkovo, Russia}

\author{A.~S.~Gorodnitskiy}
\affiliation{Russian Quantum Center, 143025, Skolkovo, Russia}
\affiliation{Moscow Institute of Physics and Technology, 141700, Dolgoprudny, Russia}

\author{M.~V.~Ryabko}
\affiliation{Samsung R\&D Institute Russia, SAIT-Russia Laboratory, 127018, Moscow, Russia}

\author{S.~V.~Polonsky}
\affiliation{Samsung R\&D Institute Russia, SAIT-Russia Laboratory, 127018, Moscow, Russia}

\author{M.~L.~Gorodetsky}
\email[]{mg@rqc.ru}

\affiliation{Russian Quantum Center, 143025, Skolkovo, Russia}

\affiliation{Faculty of Physics, M.V. Lomonosov Moscow State University, 119991
Moscow, Russia}

\maketitle

\noindent\textbf{\noindent
Narrow linewidth lasers and optical frequency combs generated with mode-locked lasers have revolutionised optical frequency metrology. The advent of soliton Kerr frequency combs in compact crystalline or integrated ring optical microresonators has opened new horizons in academic research and industrial applications. These combs, as was naturally assumed, however, require narrow linewidth single-frequency pump lasers. We demonstrate that an ordinary cost-effective broadband hundreds milliwatts level Fabry--Perot (FP) laser diode, self-injection locked to a microresonator, can be efficiently transformed to a powerful single--frequency ultra--narrow linewidth light source with further transformation to a coherent soliton comb oscillator. Our findings pave the way to the most compact and inexpensive highly coherent lasers, frequency comb sources, and comb-based devices for mass production.
}

Kerr optical frequency combs in high-Q optical microresonators\cite{DelHaye2007} are attracting growing interest in recent years\cite{Diddams2011,Maleki2016}, especially since the mode-locking via generation of dissipative Kerr solitons (DKS) has been demonstrated on a variety of platforms\cite{KippenbergScience, Kippenberg2014,Kippenberg2016,Vahala2015,Maleki2015,Gaeta2016,Weiner2016}. DKS have enabled compact and broadband low-noise frequency combs with repetition rates in the multi-GHz to the THz domain. DKS have been applied for dual comb spectroscopy\cite{Vahala2016, dutt2018chip, yu2018dualmidir}, coherent communication\cite{marinpalomo2017communication, Fulop:17}, ultra-fast ranging\cite{Suh2017,Trocha2017}, low-noise microwave master oscillators\cite{Maleki2015}, calibration of astronomical spectrometers\cite{obrzud2017astrocomb, suh2018searching} and imaging of soliton dynamics\cite{Lucas_18, yi2018imaging}. Integration of high-Q microresonators suitable for soliton generation has advanced significantly\cite{moss2013cmosplatform, yang2018bridging, Arafin}. A recent breakthrough was the assembly of an integrated photonics based optical frequency synthesiser\cite{DODOS2017} pumped with an external III-V/silicon based laser\cite{liang2010recent}. Yet, most of these demonstrations used single frequency lasers with amplifiers and modulators for soliton generation restricting commercialisation, e.g. highly sensitive wearable spectrometers and ranging sensors. 

A straightforward approach to obtain DKS in microresonators uses single frequency narrow linewidth tunable lasers for pumping. The frequency of the laser, having a linewidth comparable with the width of microresonator resonance, is precisely tuned to a mode frequency with a required red-detuned offset for a soliton frequency comb generation\cite{Kippenberg2014}. However, bulky fiber lasers or external cavity stabilised diode lasers do not naturally fit tiny microresonators.

The industry is waiting for cost-effective narrow-linewidth lasers required for pumping microresonators and further optical frequency comb generation and other applications. Near future commercialisations include optical communications with extremely high data transfer rates for data centres\cite{Soma2018}, optical sensing for wearables\cite{White2007}, light detection and ranging (LIDAR) for rapidly growing advanced driver-assistance systems (ADAS)\cite{Dale:14}. Future industry applications may include dual-comb Coherent Anti-Stocks Raman Scattering (CARS)\cite{Lai2011}, atomic clocks for global positioning \cite{Bloom2014} and quantum optics\cite{Prtljaga2016}. With compact and powerful highly coherent light source, one can realise a chip-scale nanophotonic coherent 3D imager based on frequency modulated continues wave (FMCW) LIDAR\cite{Aflatouni:15}.


\begin{figure*}[ht]
	\includegraphics[width=1\textwidth]{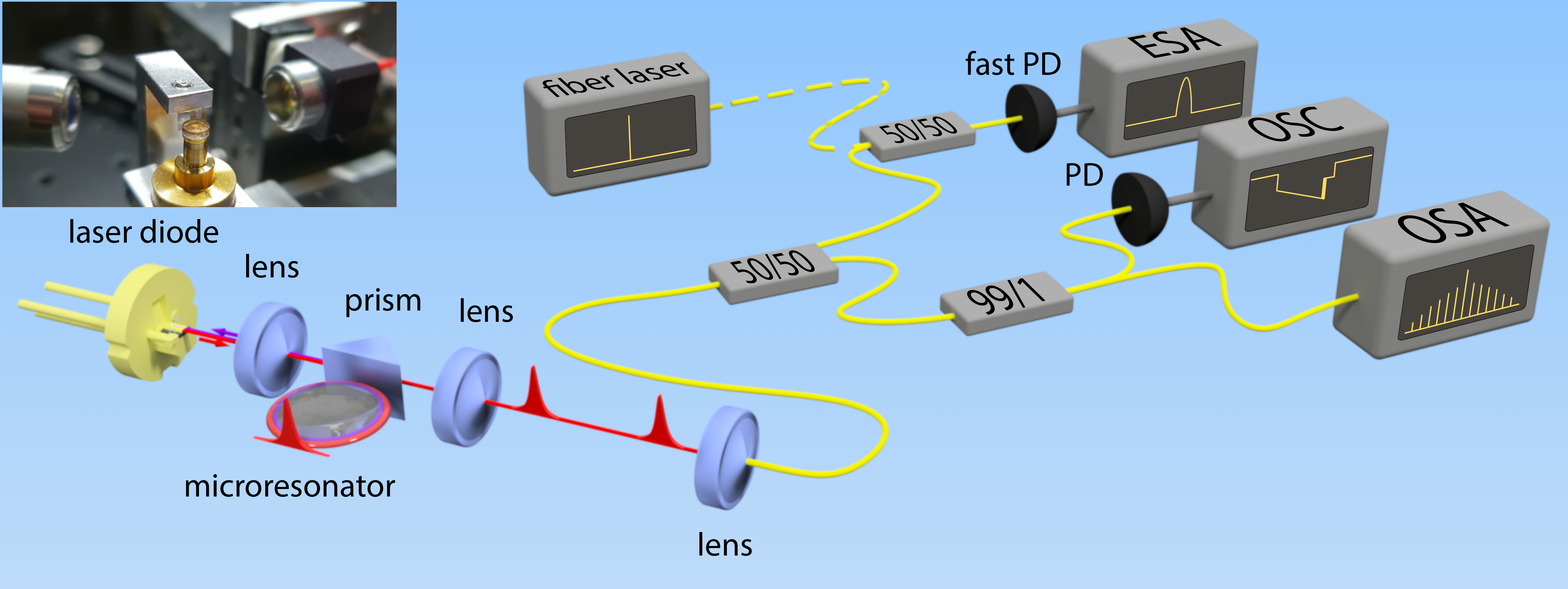}
	\caption{\textbf{Experimental scheme and set-up.} An uncapped laser diode excites a WGM microresonator using a collimator and a prism coupler. Depending on the regime, either narrow linewidth CW lasing or soliton pulses corresponding to coherent frequency comb are observed. For the linewidth measurement of the self-injection locked laser, an additional narrow linewidth laser was used for heterodyning. PD: photodetector; OSC: oscilloscope; ESA: electrical spectrum analyzer; OSA: optical spectrum analyzer.}
	\label{ris:image1}
\end{figure*}

\section*{Self-injection locking with microresonators and microcombs}
The solution proposed in\cite{Maleki2015} for a compact microresonator based Kerr comb used a relatively broadband (10-1000 times wider than the microresonator's resonance) single-frequency distributed feedbacak (DFB) laser. In this case, the role of the microresonator was twofold:
1) acting as an external cavity it narrowed the linewidth of the laser via the self-injection locking effect\cite{Vasil'ev1996,Yarovitsky1998,Donvalkar2018}, and 2) it provided a nonlinear low-threshold Kerr medium where a frequency comb appeared if the laser was appropriately detuned. Laser linewidth narrowing in this approach exploits the coupling of a free-running laser diode with the WGM microresonator\cite{Braginsky1989} having internal and surface Rayleigh backscattering\cite{Ilchenko1992,Ilchenko2000}. These scattering effects resonantly couple an excited traveling wave WGM with the identical but counter propagating mode which returns back to the laser, locking it to the cavity resonance. This technique was used for stabilisation of a DFB laser down to a record sub--Hz level\cite{Liang2015}. The analytical theory of self-injection locking by a WGM cavity initially proposed in\cite{Oraevsky} was recently revised and extended in\cite{Kondratiev2017}. 

It was previously assumed for granted that `only single-mode DFB lasers characterized with comparably high internal Q's are suitable for stable self-injection locking using multimode optical cavities'\cite{Xie:15}. Similarly, earlier only single-frequency pre-stabilized external cavity diode lasers (ECDLs) with diffraction grating\cite{Yarovitsky1998, Velichansky2003} were considered for self-injection locking to WGM cavity\cite{Maleki2015, Maleki2010}. Any external cavity pre-stabilisation complicates devices and their integration. The DFB lasers, however, are not available for many wavelengths and have a limited power that in turn restricts the power of single-mode operation and generated frequency combs at milliwatt level. Meanwhile many applications require higher comb power. For example, absorption dual-comb spectroscopy using surface diffuse scattering has less than $1$\% of light collection efficiency so that high power of incident combs must be provided\cite{Hensley}. Evaluations in\cite{Shchekin:18} show that dual-comb CARS spectroscopy of blood glucose provides a measurable glucose signal at about 100mW power of frequency combs with $10$~GHz mode spacing.

\section*{Single-frequency lasing with a multi-frequency laser}

We revealed that the initial mode pre-selection and pre-stabilisation in laser diodes are not required to obtain stable narrow-linewidth single-frequency lasing, and a WGM microresonator can handle all of these purposes efficiently as well. Consequently, simpler and cheaper FP laser diodes with higher power may be used.  

We demonstrate for the first time an efficient (up to 50\%) conversion of a broadband multifrequency FP laser diode, coupled to a high-Q WGM microresonator, into a narrow linewidth single-frequency light source in the 100 mW power range at optical telecom wavelength, with its subsequent transformation to a single-soliton Kerr comb oscillator. FP laser diode spectrum narrowing occurs in regime of competition between many longitudinal modes. Self-injection locking solves two critical technical problems of soliton Kerr combs: (1) thermal instability and (2) preferential excitation of multiple-soliton regimes. The soliton states are only possible at strong red detunings from the cavity resonance where the CW internal circulating power is small. That is why, the transition from a chaotic comb to a multiple solitons comb and finally to the single-soliton state via slow laser tuning leads to a fast drop of the internal power resulting in cooling of the resonator and finally, due to thermal refraction and expansion effects, to a large detuning from the required regime causing the loss of the soliton state. Detuning can be compensated by an electronic feedback which requires fast tuning (with the characteristic time of thermal relaxation of the resonator) of the laser frequency which is difficult to achieve. That is why different additional nonstationary `power kicking' methods\cite{Kippenberg2016, Brasch2016, Yi2016} were proposed to reach the thermal equilibrium. However, the optical feedback in self-injection locking is fast enough to compensate thermal effects in real time. Additionally, slower tuning from a chaotic to soliton state, only possible with the supported thermal equilibrium, allows a transition to smaller initial numbers of circulating solitons and finally to a single-soliton state\cite{Lobanov2016}.

\section*{Experimental setup} 

\begin{figure*}[ht]
	\centering
	\includegraphics[width=1\textwidth]{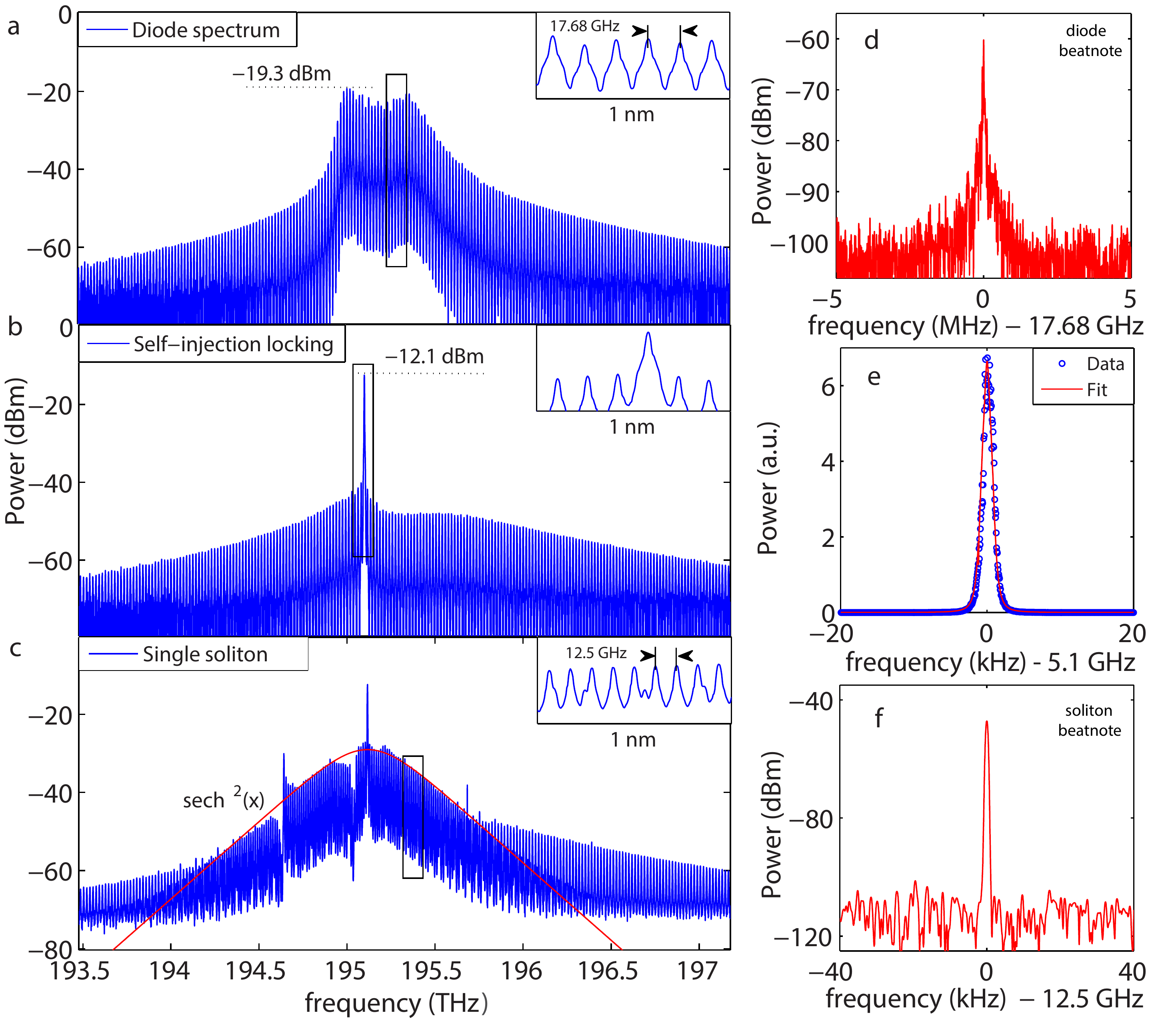}
	\caption{\textbf{Self-injection locking and spectral narrowing of a multi-frequency diode laser coupled to a MgF$_2$ ultra-high-Q whispering gallery microresonator.} \textbf{(a)} The spectrum of the free-running diode laser. \textbf{(b)} The spectrum of the diode laser stabilised by a high-Q microresonator. \textbf{(c)} Soliton generation in self-injection locking regime. \textbf{(d)} The FSR beat note signal of the free running multi-frequency laser, the beat note frequency corresponds to the diode chip length of 2500 $\mu m$. \textbf{(e)} Heterodyne signal between self-injection locked diode laser and narrow linewidth fiber laser -- blue curve, Voigt fit -- red curve. \textbf{(f)} repetition rate signal of a single-soliton state, central frequency corresponds to a WGM cavity with 5.5 mm diameter (ESA RBW is $1$~kHz).}
	\label{ris:image2}
\end{figure*}

A schematic view and a picture of the experimental setup is presented in Fig.\ref{ris:image1}. 
The laser beam from a free-space multi-frequency InP diode is collimated and coupled to a MgF$_2$ WGM resonator with a glass (BK-7) prism \cite{Gorodetsky1999}. Resonantly backscattered Rayleigh radiation returns to the diode laser and forces self-injection locking of the laser frequency to the microresonator's WGM mode. The output beam is collimated to a single-mode fiber and analyzed with an optical spectrum analyzer (OSA), on a photodiode (PD) with an oscilloscope (OSC), and an electrical spectrum analyzer (ESA). The repetition rate of the soliton pulses is monitored by a fast photodiode and ESA. The detuning of the laser frequency from an optical resonance is monitored on a PD with an oscilloscope. A narrow linewidth tunable fiber laser is used for the heterodyne linewidth measurements.

For pumping millimeter-sized MgF$_2$ resonators, ordinary packaged uncapped free-space multi-frequency laser diodes were used (Seminex, chip length $L=2500$~$\mu$m, central wavelengths 1535~nm, 1550~nm and 1650~nm covering spectral intervals of $\Delta \lambda \sim 10$~nm and a total power of $P\sim$200~mW). Generation of the self-injection locked soliton combs with a repetition rate signal linewidth $\sim1$~kHz was observed when the laser diode driving current was manually  adjusted to red detune the pump frequency in self-injection locked regime within a soliton-supporting high-Q cavity resonance.

The experimental results demonstrated in Fig.\ref{ris:image2} were obtained with a MgF$_2$ resonator with a diameter of $5.5$~mm and edge curvature radius of $500 \mu$m corresponding to the free spectral range (FSR) of $\sim 12.5$~GHz (inverse of the pulse round-trip time in the microresonator). The group velocity dispersion (GVD) for all tested laser frequencies is anomalous allowing for the generation of DKS. The microresonator was manufactured by precise single-point diamond turning (DAC ALM lathe, see SI)\cite{Tanabe2016}. The ultra-high intrinsic Q-factor exceeding $10^9$ was achieved by polishing with diamond slurries\cite{Maleki2001}. Experimental results with other microresonators pumped at different wavelength are presented in SI.

The laser diode used to obtain results in Fig.\ref{ris:image2}(a) has an optical spectrum consisting of tens of incoherent lines covering $\sim 10$~nm with mode spacing $\Delta f=\frac{c}{2Ln} \approx 17.68$~GHz around the central wavelength $1535$~nm and $200$~mW maximum output power. The intensity in the laser gain region is approximately uniformly distributed between the lines. Fig.\ref{ris:image2}(d) shows the noisy beatnote signal from the adjacent lines of the free-running laser diode at $17.68$~GHz with $\sim 1$~MHz full width at half maximum (FWHM) linewidth. The light back-reflected from the microresonator due to resonant Rayleigh backscattering, provided in case of crystalline materials mostly by surface inhomogeneities, leads to natural feedback for the laser diode. Back reflection is measured in our case at $10^{-3}$ intensity level (see SI for details). The backscattering intensity depends on the degree of loading (coupling efficiency) of the resonator, and thus can be regulated by changing the gap between the prism and the resonator\cite{Yarovitsky1998}. 

\section*{Self-injection locking with a multi-frequency laser diode}

Fig.\ref{ris:image2}(b) demonstrates a collapse of the wide spectrum of the multi-frequency laser diode to a single-frequency line. In case of a multi-frequency diode in the self-injection locking regime, the power from multiple modes due to mode competition is transferred into a single narrow line, and its output power increases. In this way, the microresonator behaves not like a simple filter cavity but plays an active role in lasing. Fig.\ref{ris:image2}(b) illustrates that in the case of self-injection locking to the WGM microresonator, the power in the dominant line increases by $\sim 7$~dBm. This effect gives a significant additional advantage of using longer multi-frequency FP diode chips with higher power as compared to DFB lasers (with maximal power $\sim 40$~mW at telecom wavelengths) \cite{Maleki2010,Maleki2015}. The asymmetry of residual laser modes in the self-injection locking regime in Fig.\ref{ris:image2}(b), with the dip on the high-frequency wing, is associated with the anomalous interaction of spectral modes in a semiconductor laser\cite{Bogatov1975, ahmed2002,AhmedYamada2010}. Note that the contrast between the dominant line and residual lines of an order of 35~dB may in our case be significantly improved with an additional coupler prism as a drop port. In this case, the residual modes of the laser will be filtered out by the resonator. Recently a 446.5 nm  self-injection locked laser \cite{Donvalkar2018} was demonstrated with sub-MHz linewidth by using a high-${Q}$ (${Q>10^9}$) WGM ${\rm MgF_2}$ microresonator in conjunction with a multi-longitudinal-mode laser diode. The presented blue FP laser had a two peak spectrum at low driving current in the free running regime. Longitudinal mode competition and conversion efficiency of the laser power to the power of a single mode were not considered.  

We used the heterodyne technique to measure the instantaneous laser linewidth in the self-injection locking regime operating at $1550$~nm. The beatnote of the self-injection locked diode laser with the narrow linewidth tunable fiber laser (Koheras Adjustik) was analyzed on a fast PD and is shown in Fig.\ref{ris:image2}(e). The Voigt profile\cite{Stephan2005} fit provides a Lorentzian width contribution due to laser white frequency noise of $370$Hz, and Gaussian contribution due to flicker frequency noise of $1.7$ kHz. The theory of self-injection locking developed in\cite{Kondratiev2017} allows  this linewidth to be estimated (see the Supplementary Information (SI) for details) with a good agreement.  In this way, an efficient and compact single-mode diode laser with a narrow linewidth at kHz range is demonstrated.

\section*{Soliton microcomb with a multifrequency laser}

\begin{figure*}[ht]
	\centering
	\includegraphics[width=1\textwidth]{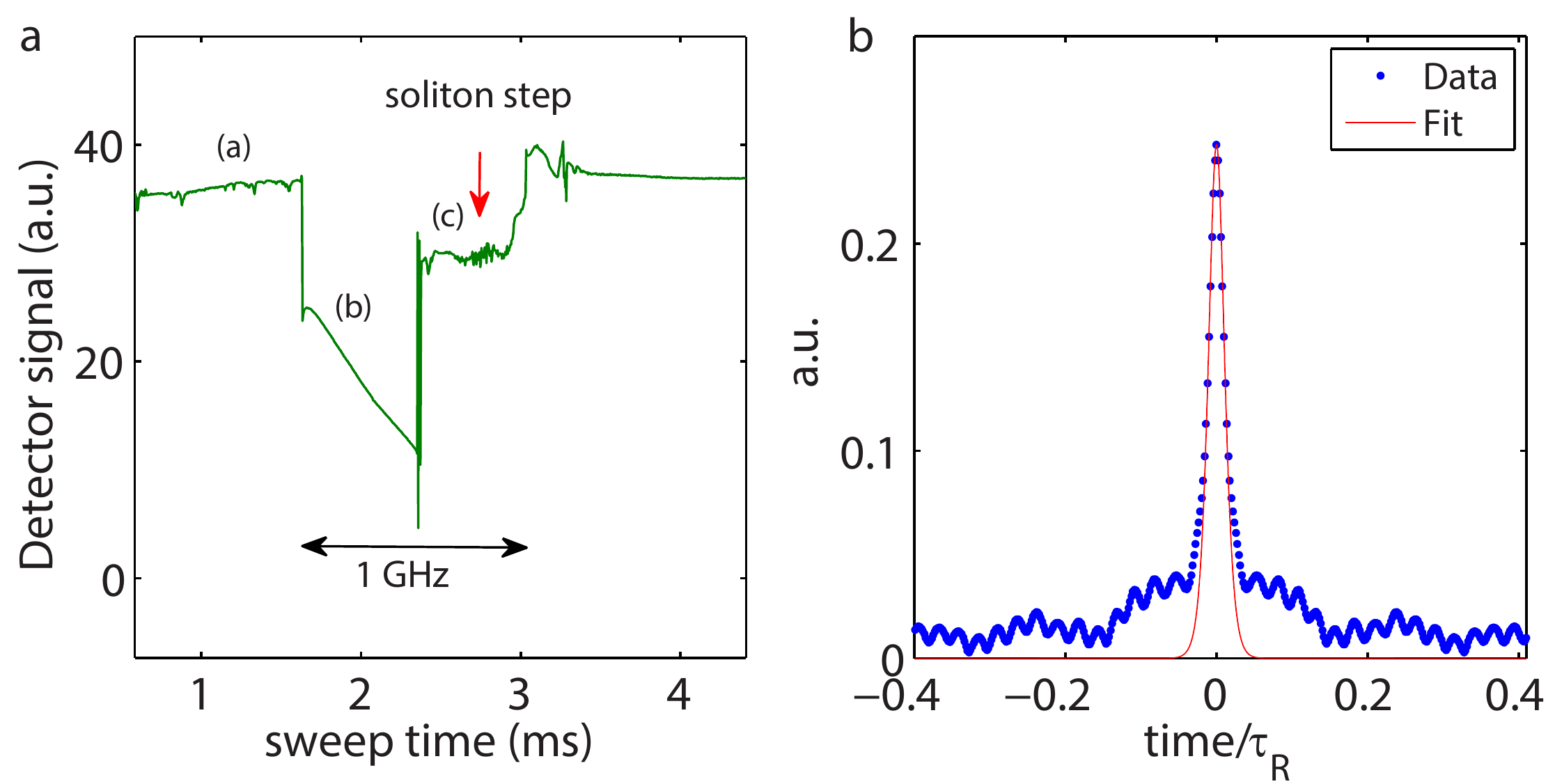}
	\caption{\textbf{Soliton generation with a laser diode.} \textbf{(a)} Cavity mode response on oscilloscope when the frequency of the laser is swept through the WGM cavity resonance, characteristic for the self-injection locking, with step-like transition to a soliton state. \textbf{(b)} The autocorrelation of a soliton waveform obtained from the spectrum in Fig.\ref{ris:image2}b (blue) and the theoretical fit (red). $\tau_R = 80 ps$ is the roundtrip time.}. 
	\label{ris:image3}
\end{figure*}

Fig.\ref{ris:image3}(a) illustrates the typical cavity response on PD when the frequency of the laser is swept through the WGM cavity resonance, characteristic for the self-injection locking\cite{Kondratiev2017}, with step-like transition to a soliton state. The self-injection locking range, calibrated with a fiber-loop cavity, was of an order of $1$~GHz, depending on the quality-factor of the WGM resonance. Region (a) corresponds to the free-running laser (Fig.\ref{ris:image2}(a)), region (b) corresponds to the case of a self-injection locking regime without a frequency comb (Fig.\ref{ris:image2}(b)), and region (c) is a soliton existence region analogous to the soliton step obtained in experiments with tunable single frequency lasers\cite{Kippenberg2014} (Fig.\ref{ris:image2}(c)). We use MgF$_2$ material for microresonator fabrication where thermo-optical instability is suppressed due to the same sign of thermo-refractive and thermal expansion effects and stable soliton generation is possible. In the self-injection locking regime, if the characteristic time of back-reflection is shorter than the thermal relaxation time, the laser frequency follows the cavity thermally shifted resonance and thermo-optical instabilities are suppressed. We observed partial suppressing of thermal nonlinearity while scanning the laser across the cavity resonance (Fig.\ref{ris:image3}(a)) due to self-injection locking and we were able to tune into stable regime of soliton generation without thermal jumps.

By gradually red detuning the diode laser frequency with the current, but staying in the locked regime, we could smoothly switch the system into a soliton comb regime (predominantly single-soliton one) with a very characteristic sech$^2$(x) envelope Fig.\ref{ris:image2}(c). The soliton  comb has a span $\sim 30$~nm with a line spacing of $12.5$~GHz. Additional residual laser lines separated by $17.68$~GHz are visible in optical spectrum, although they are weak and could be filtered out in drop-port configuration. By stronger variation of the diode current it was also possible to switch to a different resonance of the microresonator, jumping from one resonance to another within a single mode family, thus gradually changing the central frequency of the soliton and its bandwidth (see SI Fig.S.6). 

The soliton repetition rate signal in the microwave range is demonstrated in Fig.\ref{ris:image2}(f). Fig.\ref{ris:image3}(b) shows the result of the inverse Fourier transform of the spectrum from \ref{ris:image2}(c) and a fit $2A^2t/\sinh(Bt)$ which corresponds to the soliton waveform $A{\,\rm sech}(Bt)$, which reveals the soliton duration of $220$~fs. The residual fringes are caused by a dispersive wave formed due to parasitic mode-crossings which one can see in Fig. \ref{ris:image2}(c). 

Note, that no particular technique with amplitude and frequency manipulations and hence no additional equipment was used to control the soliton generation in the presence of thermal nonlinearities\cite{Kippenberg2014}. This is possible due to the very fast optical feedback which compensates thermal cavity detuning upon switching. Single-soliton states lived several hours in laboratory conditions without any additional stabilisation techniques -- another convenient consequence of self-injection locking. In our experiment coupling rate to soliton resonances was around 15--25\% due to the non-optimised geometry of the microresonator and prism coupling\cite{bilenko2017optimisation68043753}, so we did not efficiently use available laser power to generate a wide frequency comb. Nevertheless, our result shows that this proposed method is applicable due to pump power overhead even when coupling conditions are not optimal or the Q-factor is not ultra high, e.g. for integrated microresonators.

We have checked that the Kerr soliton comb is generated inside the microresonator and is not generated or amplified in the FP laser chip by adding a beam splitter between the laser chip and the microresonator and by observing a spectrum of the light immediately after the gain chip. In this way we confirm that only single frequency lasing (corresponding to  Fig.\ref{ris:image2}(b)) is observed at the output facet of the laser. It should be noted that in our work the FSRs of the FP diode laser and microresonator did not match.

We observe that single soliton generation is preferable although multi-soliton states are also possible. This results from a relatively slow transition from CW to soliton regime \cite{Lobanov2016}, due to low speed of pump frequency tuning in the locked regime. Comparing to previous realisations of tuning methods to obtain soliton states (fast forward scan\cite{Kippenberg2014}, slow backward tuning\cite{Karpov16}, pump power kicking\cite{Brasch2016}), in self-injection locking the frequency tuning is orders of magnitude slower. Considering realistic parameters of our system we estimate the tuning speed is $10^4$ times slower than without self-injection locking\cite{Kondratiev2017} (see SI for details)

We have also checked and confirmed the generation of the soliton comb in the same microresonator using the traditional technique with a CW narrow linewidth tunable fiber laser (see SI for details). 

\section*{Discussion and conclusion}

In conclusion, we have demonstrated a new efficient method for achieving a single-frequency narrow linewidth lasing and independently a method for generating stable Kerr soliton combs directly from multi-frequency high-gain laser diode chips using the self-injection locking effect. This result paves a way to compact, low-noise photonic microwave sources and for generating stable powerful frequency combs, which are important for spectroscopy, LIDAR application, astronomy, metrology and telecommunications.  

\section*{Methods}

{In experiments we used for the self-injection locking with high-Q MgF$_2$ WGM resonators Indium Phosphide multifrequency and single latitudinal mode laser diode chips. The length of chips was 1.5--2.5 mm and the free running spectrum consisted approximately of 50 FP lines with a beatnote between adjacent diode FP lines 1--3~MHz. The output power was 100--500~mW depending of the diode length and applied current. Such chips are commercially available in wide wavelength range.}




\section*{Data availability statement}
All data used in this study are available from the corresponding authors upon reasonable request.

\section*{Authors contributions}
Experiments were conceived by N.G.P., S.K., and M.L.G. Analysis of results was conducted by N.G.P., A.S.V, S.K., G.V.L. and M.L.G.,  N.G.P., A.S.V., S.K. and A.S.G performed measurements with diode lasers and G.V.L. with a fiber laser. G.V.L., N.G.P. and A.S.V. fabricated devices. S.V.P. and M.R. set the research direction relevant for industrial needs - comb source for wearable spectrometer (Samsung Gear). S.V.P. supervised the project from the Samsung. M.L.G. supervised the project. All authors participated in writing the manuscript.

\begin{acknowledgments}
This publication was supported by the Russian Science Foundation (17-12-01413). G.V.L., N.G.P. and A.S.V. were partially supported by the Samsung Research center in Moscow.  The authors gratefully acknowledge valuable discussions with Tobias Kippenberg, Kerry Vahala and Vitaly Vassiliev. The authors express gratefulness to Hong-Seok Lee and Young-Geun Roh from Samsung Advanced Institute of Technologies for help in establishing the project and its further support.
\end{acknowledgments}

\clearpage

\end{document}